\documentclass[twocolumn]{aastex631}
\usepackage{amsmath}

\begin{document}

\title{Clump-fed black hole growth in the first billion years of the universe}

\author[0009-0008-2970-9845]{Manish Kataria}\thanks{Email: manish@iucaa.in}
\affiliation{Inter-University Centre for Astronomy and Astrophysics, PostBag 4, Ganeshkhind, Pune-411007, India}

\author[0000-0002-8768-9298]{Kanak Saha}\thanks{Email: kanak@iucaa.in}
\affiliation{Inter-University Centre for Astronomy and Astrophysics, PostBag 4, Ganeshkhind, Pune-411007, India}

\author[0000-0002-1723-6330]{Bruce Elmegreen}
\affiliation{Katonah, NY 10536, USA}

\begin{abstract}

Understanding how supermassive black holes (SMBHs) form in the early universe is one of the most challenging problems in astrophysics. Their high abundance in the first billion years, as observed by the James Webb Space Telescope, hints towards black hole seeds that accrete mass rapidly. The origin of this accreted mass is not known. Here, we consider a billion solar mass clumpy galaxy at z=5.48 with a 30 million solar mass black hole in the center. We show that the clumps should migrate to the central region because of torques from dynamical friction with the halo, funneling in at least $14$ solar masses per year. This is fast enough to grow the observed SMBH, with only 1\% of the accreted mass getting in and the rest going to a bulge. Clump-fed accretion could explain most young SMBHs because young galaxies are highly irregular with massive star-forming clumps.

\end{abstract}


\section{Introduction} \label{sec:intro}

High-redshift star-forming galaxies are characterized by clumpy and irregular structures which arise due to gravitational instabilities in their turbulent, gas-rich disks \citep{Cowie1995, vandenbergh1996, Conselice2004, Elmegreen2005, Genzel2006, Bournaud2007, Dekel2009, Mandelker2014, Kalita2025a}. Each clump, also sites of star-formation, in these high-z galaxies accounts for a significant percent of the galaxy's stellar mass \citep{ElmegreenDebra2009, Guo2018, Kalita2025b}. Supermassive black holes (SMBHs) within these galaxies are observed to possess masses ranging from $10^{6}\ \rm{to}\ 10^{9}M_{\odot}$ \citep{Heckman2014} even when the universe was less than a billion years old. The mechanisms enabling such rapid SMBH growth during these early epochs remain a subject of active investigation in galaxy evolution. While SMBHs in local galaxies, such as the Milky Way, represent only a small fraction of the total stellar mass (~0.01\%; \cite{Ghezetal2005}), a well-established correlation between galaxy bulge properties and SMBH mass suggests a co-evolutionary relationship\citep{Kormendy&Ho2013, Yang2019}. 

SMBH growth models suffer from significant uncertainties regarding the initial seed mass. Two primary seed formation pathways have been proposed: heavy and light seed. Heavy seed requires scenarios like the direct collapse of gas (DCBH) \citep{Begelman2006, Wise2008, Regan&Haehnelt2009, Shang2010, Johnson2011, Agarwal2013, Visbal&Haiman2018, Pacucci2023, Bogdan2024} or via the stellar collision in a dense star cluster \citep{Devecchi&Volonteri2009, Devecchi2010, Davies2011}. On the other hand, the light seeds can be formed as remnants of Pop III stars, the first generation of stars. These massive black holes can have a mass of about a $100M_{\odot}$ \citep{Fryer&Kalogera2001, Madau&Rees2001, Bromm2009}. Discriminating between these two scenarios in galaxies at redshifts $z \lesssim 8$ is challenging because their luminosity functions become indistinguishable over time due to accretion and growth \citep{Ricarte&Natarajan2018}. Recent studies with JWST provided some evidence in favor of the heavy seeding scenario specifically due to the presence of massive black hole masses of  $\sim 10^{6} - 10^{7}M_{\odot}$ at z$\sim$10 \citep{Maiolino2024, Bogdan2024, Kovacs2024}. However, there are alternate explanations put forward as well - that these detected SMBH could be the tail end (i.e., the most massive members) of the distribution, with major populations consisting of lower masses ($M_{BH} \lesssim 10^{5}M_{\odot}$) being undetected due to the sensitivity of the JWST and the selection effects \citep{Li2024}. 

Given the assumption that lower mass black holes, which are undetected, at high-z form a major population of the distribution around the peak, the light seed origin is favored. These light seeds need to grow rapidly compared to the heavy seeds to explain the observed population of black holes early in the universe. Several pathways have been proposed to explain the growth mechanism in support of the light seed origin, e.g., smooth Eddington accretion with small bursts of super-Eddington \citep{Volonteri&Rees2005, Madau2014, Trinca2024}, growth via black-hole mergers \citep{Micic2007}, and galaxy merger-driven accretion \citep{Hopkins2006}. These mechanisms collectively explain the emergence of SMBH within a billion years after the Big Bang. These models have mainly been evaluated on their ability to reproduce the observables like AGN luminosity function and the SMBH galaxy scaling relations observed in the nearby universe. Although it is hypothesized that frequent galaxy mergers can drive the SMBH growth at high-z, some studies also point towards the secular growth of the galaxy-SMBH relations with merger not being a dominant mechanism \citep{Cisternas2011, Kocevski2012}. \cite{Smethurst2022} produced the scaling relations in the merger-free population of galaxies in simulation. They showed that mergers help reduce scattering in correlations; otherwise, secular evolution can also give rise to scaling relations.

In the high-z universe (when most galaxies are clumpy), there have been efforts to explain the observed scaling relations and black hole growth mechanism via the accretion of massive clumps as an alternative channel \citep{Bournaud2011}. Numerical simulations by \cite{Elmegreen2008b} demonstrate that intermediate-mass black holes (IMBHs), formed within dense young star clusters, can efficiently migrate toward the galaxy center via the influence of dynamical friction. In the end, these processes can produce the observed SMBH-Bulge scaling relations. Subsequently, using high-resolution simulations \cite{DeGrafetal2017} proposed the concept of clump-assisted growth of the SMBH at the center of high-z galaxies due to the accretion of material from the inspiraling clumps. \cite{DeGrafetal2017} showed that these accretions are efficient in rapidly growing the SMBHs, while complementing the smooth accretion mechanism. Motivated by the findings from analytical and simulation-based studies, we test the clump-assisted black hole growth hypothesis on publicly available observations from HST, JWST, and VLT/MUSE. Our analysis focuses on a galaxy at z = 5.48 (hereafter GSz5BH), first identified by \cite{Rhoads2005} as a tadpole galaxy in the Hubble Ultra Deep Field \citep{Beckwith2006}. HST/G141 Grism data confirmed the observed clumps in the images are part of the GSz5BH with Ly$\alpha$ emission thanks to high spatial resolution. High redshift, clumpy nature, and the presence of AGN \citep{Matthee2024} made GSz5BH an ideal candidate to study the clump-fed growth mechanism with the highest resolution UV-Optical-NIR restframe imaging and IFU data covering the Ly$\alpha$ regime.

This paper is organized as follows: Section 2 outlines the data and source selection, while Section 3 focuses on extracting the photometry and modeling the broad-band SED of the full galaxy and the individual clumps,  with the removal of AGN light contribution via modeling the PSF of each HST and JWST band. Section 4 details the analysis of MUSE data and the kinematic measurements. Section 5 deals with the analytical calculation of black hole growth and clump accretion timescales. 

A flat $\rm \Lambda$~CDM cosmology with $\rm H_{0}$ = 70 km s$^{-1}$ Mpc$^{-1}$ , $\Omega_{m}$ = 0.3, and $\Omega_{\Lambda}$ = 0.7 was adopted throughout the article. All magnitudes quoted in the paper are in the AB system (\cite{Oke1983}). 
 
\begin{figure}[ht]
\begin{centering}
\rotatebox{0}{\includegraphics[width = 0.5\textwidth]{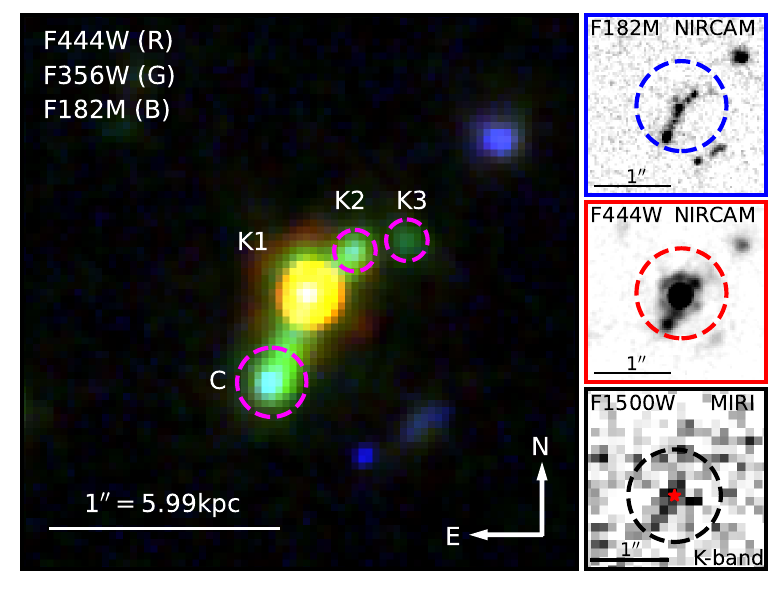}}
\vspace{-0.9cm}
\caption{\textbf{Morphology of GSz5BH}: A false-color image of the tadpole/chain galaxy at redshift 5.48 in JWST/NIRCam filters with RGB probing rest-frame Halpha~6563\text{\AA} (F444W), [OIII]5007\text{\AA} (F356W) and MgII]2799\text{\AA} (F182M) emission respectively. The galaxy has three green clumps, C, K2, and K3, marked in this image, and a central point-like clump (K1) with the characteristic JWST PSF in yellow. The blue clumps on the N-W direction and the southern side of the GSz5BH are foreground galaxies. On the right are  single-band images of the galaxy in restframe UV, Optical, and NIR (K-band).}
\label{fig:detection}
\end{centering}
\end{figure}

\begin{figure}[ht]
\nolinenumbers
\centering
\rotatebox{0}{\includegraphics[width=0.5\textwidth]{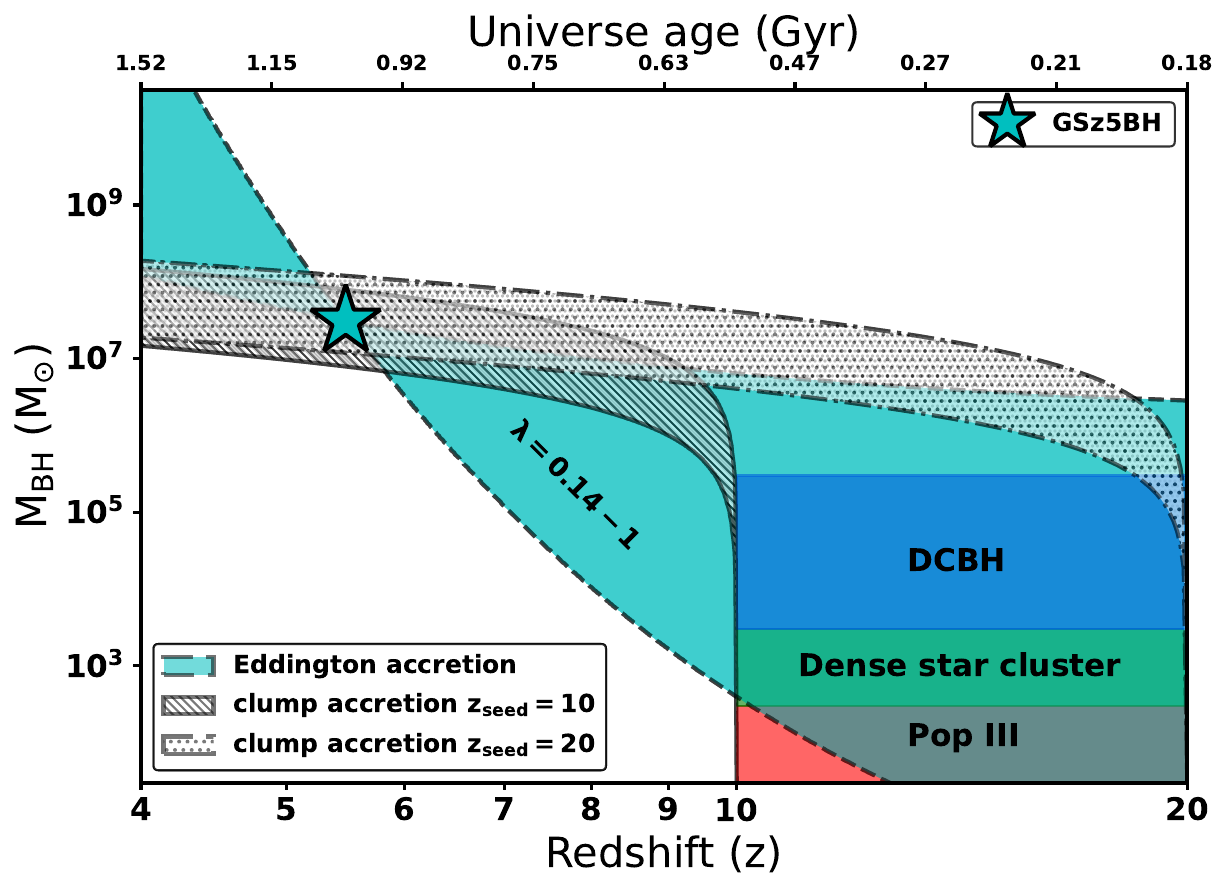}}
\caption{\textbf{BH growth: }Growth possibilities for the SMBH in GSz5BH, which is shown as a cyan star. The cyan-shaded region shows the black hole growth path following Eq. \ref{eqn:bh_growth} with Eddington ($\lambda = 1$) to sub-Eddington accretion to produce the observed SMBH mass in GSz5BH. The colored rectangles on the bottom right show the range of proposed seed masses at high-z. The hatched regions between the black lines follow Eq. \ref{eqn:clump_accretion} and show the amount of matter fed into the accretion disk via clump inspiral with $\eta$ the feeding efficiency in a range from 0.1\% to 1\%; as indicated in the lower left, the two hatched regions are for seeds starting at $z=10$ and 20.
}
\label{fig:bh_growth}
\end{figure}


\section{Data and source selection} \label{sec: source_selction}

Tadpole and chain-like elongated galaxies are ideal for studying the resolved properties of galaxies at high redshifts, primarily due to their elongated structure. These galaxies are thought to represent ongoing merging systems or edge-on disk galaxies that are settling, exhibiting clumpy morphology due to dynamic instabilities. The first identification of these tadpole/chain-like galaxies occurred in the HUDF \citep{Cowie1995, Elmegreen2005}. A sample of Tadpole galaxies was drawn from \cite{Straughn2006}, with GSz5BH listed in the catalog and also detected by their method, originating from \cite{Rhoads2005}. This galaxy represents the highest redshift tadpole galaxy identified. However, before JWST observations of the GOODS-South region, GSz5BH had not been classified as hosting an AGN.

We use publicly available high-resolution HST and JWST images in this work, which are a part of HLA (Hubble Legacy Archive) and JADES~(JWST Advanced Deep Extragalactic Survey, PI: Daniel Eisenstein \& Nora Luetzgendorf), respectively. We used the JWST/MIRI imaging of GOODS-South (PI: Rieke, George). MUSE data is taken from the MUSE HUDF observations (PI: Roland Bacon) for this study's $\rm Ly\alpha$ analyses.

\section{Photometry and SED fitting} \label{sec: photm_sed}
\subsection{Removing the AGN light from the band images} \label{subsec: agn_removal}

SED codes used to model the broadband photometry of the galaxies are based on the energy balance principle. The amount of energy of UV photons absorbed by the dust is remitted as infrared emission in mid and far IR; hence, the total energy is conserved. This principle breaks down for a spatially non-uniform distribution of the dust in a galaxy, which is the case with GSz5BH as we have more dust extinction at the location of AGN (clump K1) compared to the rest of the clumps (C, K2, and K3). To mitigate this problem, we removed the contribution of the AGN light from all broad and medium band images before performing photometric measurements.

To remove the AGN light from GSz5BH, we model it using the GALFIT \citep{Peng2002} as a PSF-like point source in all the bands. For generating input PSF, we first selected multiple isolated stars from each band image and took a cutout of these stars. We then identified the contaminating objects around the stars, mostly faint galaxies, and created a segmentation map. The contaminating objects in the segmentation map are then filled with the Gaussian noise modeled from the cutout image for each star. This process is repeated for each band used to make empherical PSF. After cleaning the star images, we stack them using the Photultils \citep{Bradley2023} python package and finally build resampled empirical PSFs using the EPSFBuilder class of the Photultils package. 

After creating the ePSFs for all the bands, we used the four times oversampled PSFs to model the point source in all the multiband images. As we can see, the PSF-like feature of the AGN only appears in the optical rest frame images significantly from JWST/NIRCam (F277W to F480M), while in the rest frame, AGN is very faint, and one does not see the PSF-like feature which is the case with all HST bands (see Fig.\ref{fig:hst_jwst_cutouts}). Therefore, to model the point source in rest-frame UV images, we constrained the center of the PSF to 3 pixels in these images using an input constraint file. While fitting the PSF in the images, we also provided the mask file to the GALFIT as an input to mask all the nearby objects after segmentation in the GSz5BH image cutouts using sep python module \citep{Barbary2016, Bertin_1996}. For the sigma files used in the GALFIT, we used the error maps provided by the JADES Survey in the case of JWST/NIRACM and weight maps from the HST images. As these images are already subtracted from the background, we did not use any sky model while running GALFIT modeling. After AGN subtraction, the images of GSz5BH retain an extended structure in the rest-frame optical images where the AGN is bright (see Fig.\ref{fig:hst_jwst_cutouts_agn_sub}).

\subsection{PSF matching and Photometry} \label{subsec: psf_matching}

After removing the AGN light from all the images, we PSF matched all the band images to the HST F160W, which has the lowest resolution among all the bands. Before matching the PSF, we reprojected the JWST/NIRCam images to the HST pixel scale, as JADES/NIRCam images have twice the spatial sampling of the HST images used in this work. For reprojecting, we use the Python reproject package, which implements the reprojection of astronomical images, assuming the given WCS information is correct. It uses the interpolation by an adaptive anti-aliased algorithm \citep{DeForest2004}. The ePSFs of all the bands are projected to the HST F160W PSF before creating the matching kernel.

The HST and JWST have geometrically different PSFs due to circular and hexagonal mirror shapes. To create a difference matching kernel between two differently shaped PSFs, we used the Pypher module \citep{Boucaud2016}. This module was developed as part of ESA's Euclid mission. Pypher is based on the Wiener filtering \citep{Wiener1950} with regularisation and considers the anisotropic nature of the PSFs to be matched, as in our case. The reprojected JWST/NIRCam and other HST PSFs are matched with the HST/WFC3\_IR F160W. We then use these difference kernels and perform a convolution between reprojected point source subtracted images in all the photometric bands. We perform the same projection and PSF convolution on the variance maps of the images to calculate the errors while measuring the flux.

After PSF homogenization of all the images, we then use SExtractor \citep{Bertin_1996} in dual image mode to measure the flux of the entire galaxy within an aperture of $\rm 1.5^{\prime\prime}$ diameter using HST F105W (not PSF matched) as a detection image to detected the entire galaxy as whole because after AGN subtraction there are disconnected pixels within the galaxy causing it to be detected as two separate sources. PSF-matched variance files are used during the flux measurement in the aperture. After calculating the flux with the help of SExtractor, we corrected the flux using the aperture correction obtained from the curve of growth of the HST F160W PSF.

After the flux measurements, we then calculated the magnitudes in each image by using $\rm m_{AB} = -2.5log(flux) + Z.P$, where Z.P is the photometric zero point of the image, for HST images zero points were given on the HLA (Hubble Legacy Archive) \citep{https://doi.org/10.17909/t91019} and for the JWST/NIRCam images we used the formula: $\rm Z.P = - 6.1 - 2.5log(PIXAR\_SR[sr/pix])$ given on the JWST calibration  \citep{https://doi.org/10.17909/8tdj-8n28}, where PIXAR\_SR value is present in the header of every JADES/NIRCam image. These magnitudes are further corrected for foreground Milky Way dust extinction from \cite{Schlafly_Finkbeiner2011}.

\subsection{Resolved photometry of clumps} \label{subsec: clump phot}

We apply the same aperture photometry method used to measure total galaxy photometry (see sec. \ref{subsec: psf_matching}) for the individual clumps (C, K2, and K3) as well using SExtractor. Unlike total galaxy photometry, we used JWST/NIRCam images only for individual clump photometry to reduce the flux contamination caused by lower spatial sampling resolution (2$\times$ less than JADES/NIRCam images) and comparatively larger PSF of HST images. We matched the PSF of all NIRCam images with the F444W filter of NIRCam. SExtractor aperture flux measurements are performed using dual image mode with an F335M medium band image as a detection image. Flux values obtained are then corrected using the F335M growth curve and corrected for foreground dust extinction.

\begin{figure*}[ht]
\centering
\includegraphics[width=7in]{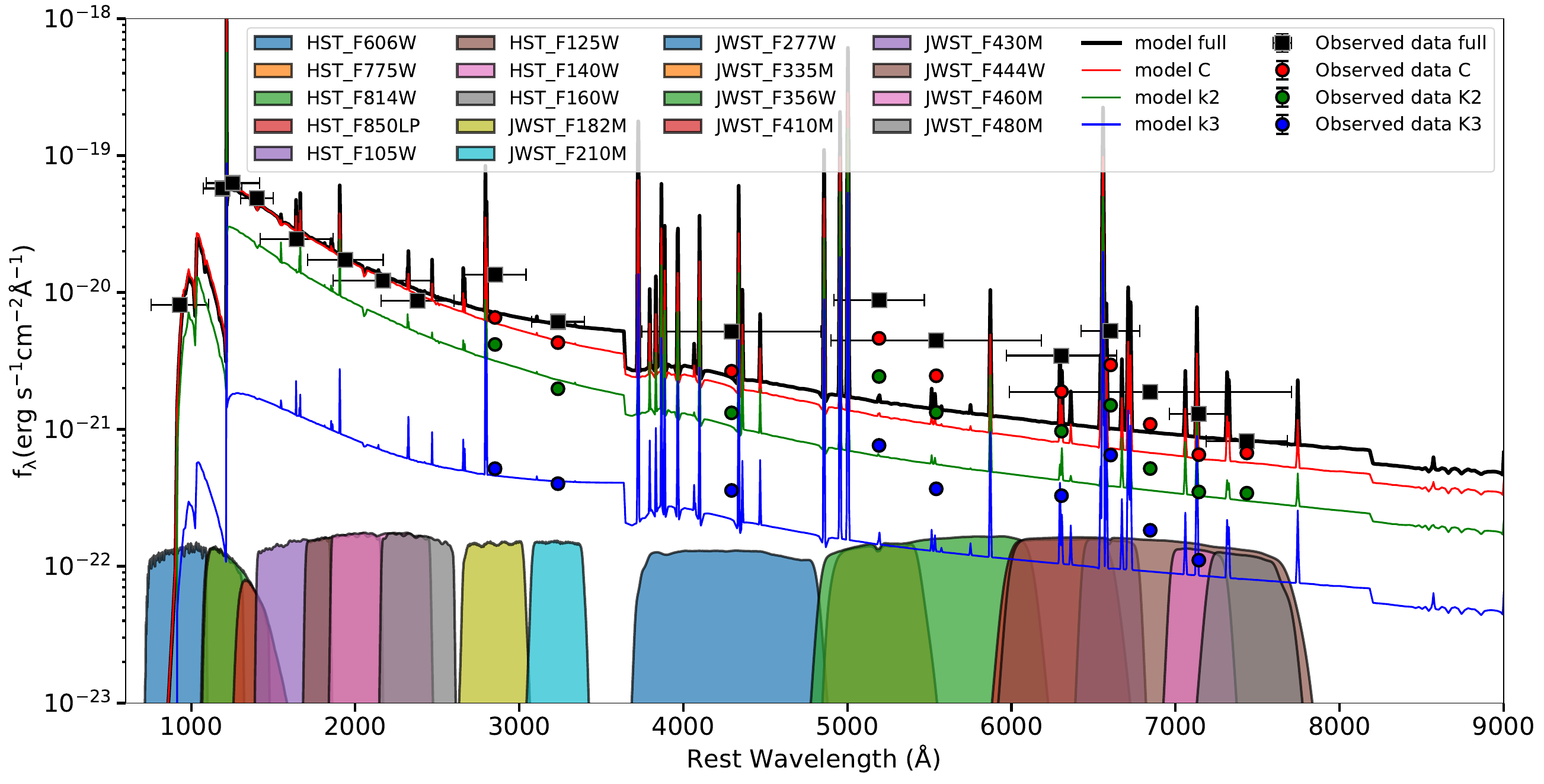}
\caption{\textbf{PSF-matched resolved SED modeling: }Model SED of GSz5BH after AGN light removal shown black the horizontal errorbar on the black square markers is the FWHM of the filters shown in color at the bottom of the plot. The red, green, and blue SEDs belong to the clumps C, K2, and K3, respectively, with data points only covering the JWST/NIRCam filters. All the models are similar in the fitted wavelength range, showing a Balmer jump, indicating a stellar population mostly consists of young stars with negligible contribution from older stars. Clump K3, being the faintest, differs in SED shape in the UV region, showing more stellar attenuation in UV.}
\label{fig:sed}
\end{figure*}

\subsection{SED Modeling}\label{subsec:sed_modeling}

We made use of a publicly available code python code CIGALE (Code Investigating GALaxy Emission) \citep{Burgarella_2005, Noll_2009, Boquien_2019} to perform SED modeling on the measured photometric data (see sec.\ref{subsec: psf_matching} and \ref{subsec: clump phot}), which doesn't contain the light contribution from AGN (see sec.\ref{subsec: agn_removal}), to estimate the stellar mass and other properties of the galaxy and its clumps. CIGALE works on the energy balance principle incorporating a dust attenuation law and re-emitting the absorbed energy in UV to mid/far IR.

CIGALE generates model template spectra of galaxies based on the combinations of various physical input parameters. For creating the stellar population, we use the double-exponential as the star formation history, sfh2exp, along with BC03 SSP \citep{Bruzual_2003} having Salpeter IMF \citep{Salpeter_1955}, which is a power law to distribute the stellar masses between 0.1-100 $\rm \rm M_{\odot}$. In the nebular module used to model the gas emission from ISM of the galaxy, we varied the ionization parameter from -2.0 to -2.6 and gas metallicity from 0.002 to 0.006 based on the previous runs. We kept the value of the LyC escape fraction between 0 and 0.2. For dust attenuation law, we use modified Calzetti law \citep{Caleztti_2000} (dustatt\_modified\_starburst, E(B-V)).

We set the E(B-V) parameter as 0, 0.03, and 0.078, where E(B-V) is the color excess or reddening to measure the dust content of the ISM in which the last two values are calculated using Ly$\rm \alpha$ and $\rm \beta-slope$, respectively. 0 value is chosen from the multiple runs of CIGALE, which always best fits. The low (or 0) value of the dust extinction in the extended parts of the galaxy (clumps C, K2, and K3) with only stellar components is also indicated by the fact that we don't observe these clumps in the MIRI images, which probes the rest frame NIR pointing that these clumps contain very young stellar population with little to no dust unlike the clump K1 hosting the AGN. The power-law slope ($\delta$), which modifies the slope of the Calzetti dust attenuation law \citep{Caleztti_2000} along with the \cite{Leitherer2002} for attenuation between 150 \text{\AA} to Lyman break, varied from -0.5 to -0.2 with an SMC-like attenuation curve appropriate for high redshift galaxies \citep{Salim2018}. For the analysis of clumps (C, K2, and K3) from only JWST/NIRCam images, We used only JWST/NIRACM images for the resolved photometry of the individual clumps as their resolutions were better than the HST after PSF matching with F160W PSF compared to PSF matching with F444W images and the two bands (F182M and F210M) covers the rest frame UV continuum as well. We followed the same procedure while doing the SED modeling of the clumps. Although we omitted the F480M filter for the K3 clump as it does not have good enough SNR in this filter but has very faint emission in the F444W filter, which is a broad-band filter covering the three medium band filters (F410M, F430M, and F480M). The resulting best-fit parameters from the SED modeling of the full galaxy (without AGN) and individual clumps are listed in Appendix Table \ref{tab:best_fit}.

\subsection{Measuring the stellar mass of the AGN obscured clump}\label{subsec:k1_mass}

To calculate the stellar mass of the clump K1, we exploit the MIRI/F1500W broad-band image of GSz5BH, which probes the K-band ($2.2\mu m$) in restframe at this redshift. We estimated the stellar mass of the total galaxy using SED modeling (see sec.\ref{subsec:sed_modeling}). These SED model fittings are after the AGN light subtraction from the galaxy, in which we also throw away the light contribution from the underlying stellar population of the clump K1 obscured by AGN. We can't measure the full galaxy mass with simultaneous AGN modeling in CIGALE due to spectral variation of AGN fraction (see Fig.\ref{fig:agn_fraction}) as CIGALE takes a single value for AGN fraction and the non-uniform distribution of dust. To measure the mass of this clump, we used V-K color, which corresponds to the F356W-F1500W at this redshift. As for the F1500W, we used JWST/MIRI images (PI: George Rieke). We subtracted the background from the MIRI image using the SEXtractor. Using the formula used for NIRCam images, we calculated the photometric zero point for F1500W (Z.P = 25.265). We correct for the systematic astrometric offset between the F356W and F1500W images using six stars and bright, compact objects (galaxies) in the image for better aperture placement for flux measurement. The calculations for stellar mass measurements are as follows.

\begin{equation} \label{eqn:mass2light}
log(M/L_{k}) = a_{k} + b_{k}\times color
\end{equation}

The values for the $a_{k} = -1.16$ and $b_{k} = 0.44$ in the equation \ref{eqn:mass2light} are taken from \cite{BelldeJong2001} for the Salpeter IMF with the lowest possible metallicity value of 0.008, the turned out to be $M/L$ = 0.139, with measured V-K color of 0.687. We calculated the luminosity for the F1500W image, and using the mass-to-light ratio, we obtained the stellar mass of the obscured clump K1, $M_{*, K1} = 2.39\times10^{8}\ \rm M_{\odot}$, which is comparable to other clump masses.

\section{MUSE Cube Analysis: \texorpdfstring{$\rm Ly\alpha\ $}\ properties}

We used the publicly available MUSE HUDF data (PI: Roland Bacon) for the $Ly\alpha$ analysis of this galaxy. This data consists of the $\sim$10 hrs of the MUSE integration time with a mean PSF-FWHM of 0.6$^{\prime\prime}$ assisted with AO.

\subsection{MUSE \texorpdfstring{Ly$\mathbf{\alpha\ }$}\ image} \label{subsec:lya_image}

Due to its extended nature, $Ly\alpha$ emission from the galaxy is best traced by the IFU data. We can define the wavelength range for constructing the narrow band image. Also, the total flux measurements are best done with IFU observations, as in the case of a slit observation, there will always be an underestimation of flux. We extracted the $Ly\alpha$ spectra of the whole galaxy and used the method described in section 3 of \cite{Drake_2017} to construct a narrow band image. We took a spectral window of 200\text{\AA} centered at the total galaxy $\mathbf{Ly\alpha}$ emission spectra and masked the emission line in the center. The rest of the spectra on both sides of the masked emission line are then used to calculate the continuum at every spaxel by fitting a constant value. The blue and reward continuum average is then subtracted from each spaxel to create an emission line cube. After subtracting the underlying continuum, we used the same emission line spectral masked window and collapsed it along the spectral axis to create a $\rm Ly\alpha$ line image of GSz5BH.

\subsection{\texorpdfstring{$Ly\alpha\ EW_{0}\ $}\ and escape fraction}

In this paper to measure Ly$\alpha$ rest equivalent width (EW$_0$) we follow the procedure described in \cite{Drake_2017} and use the following equation for EW$_0$ calculation:

\begin{equation}
EW_{0} = \frac{1}{(1+z)}\frac{F_{Ly\alpha}}{f_{\lambda,UV}}
\end{equation}

The above equation has two parts: $F_{Ly\alpha}$, which is estimated by performing a curve of growth photometry on the narrow band image of Ly$\rm \alpha$. $f_{\lambda, UV}$ is a power law in the UV (1200-3000\text{\AA}) range calculated using HST F105W, F125W and F140W bands to find the best estimates as described in \cite{Hashimoto_2017}. We found a value of rest equivalent width EW${_0}$ = 144.4 $\pm$ 6.6 $\mbox{\AA}$ which corresponds to the Ly$\alpha$ Luminosity of $\rm log(L_{Ly\alpha}\ [erg\ s^{-1}]) = 43.36$ at the redshift of this GSz5BH. We calculated the $Ly\alpha$ escape fraction, $f_{esc} = 0.693 \pm 0.203$, using the relation, $f_{esc, Ly\alpha} = 0.0048^{+0.0007}_{-0.0007}EW_{0}\pm 0.05$ from \cite{Sobral&Matthee2019}. We also calculated the $f_{esc}$ assuming $L(Ly\alpha/H\alpha) = 8.7$, which corresponds to $T = 10^{4}K$ at $n_{e} \approx 350cm^{-3}$ \citep{Hu1998}, $f_{esc} = 0.436 \pm 0.066$, using $f_{esc} = \frac{L_{Ly\alpha}}{8.7H\alpha \times 10^{0.4A_{H\alpha}}}$, where $A_{H\alpha}$ is calculated using $\beta-slope$.

\subsection{Ly\texorpdfstring{$\alpha\ $}\ halo} \label{subsec:lya_halo}

\begin{figure*}[ht]
\centering
\includegraphics[width=7in]{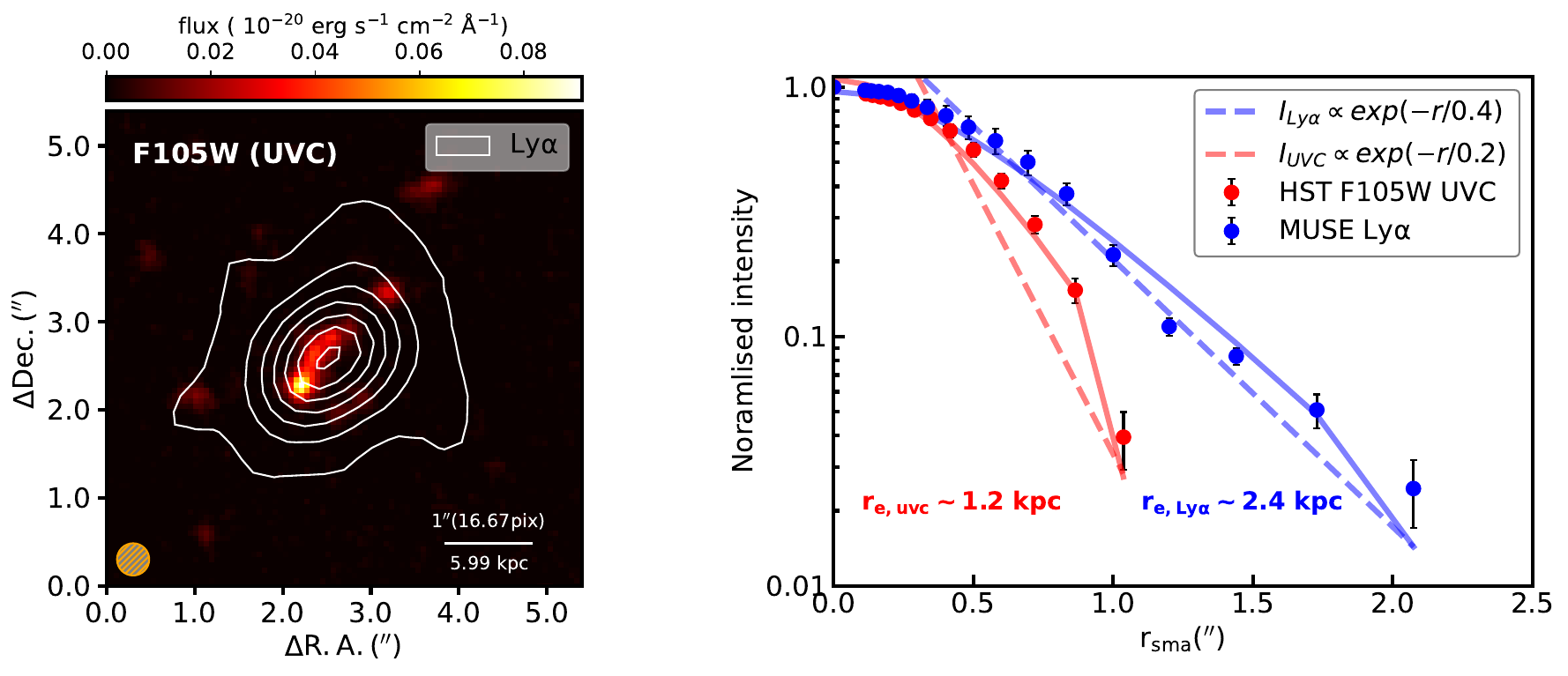}
\caption{\textbf{$\rm \mathbf{Ly\alpha}$ halo: }\textit{Left:} The image shows the restframe UV continuum of GSz5BH with the clump C being the brightest, the overplotted contours belong to $\rm Ly\alpha$ emission starting outermost being $\rm 1, 3, 5, ..., 13\sigma$, The contours are concentrated at the position of clump K1 (AGN) but also shows the elongated emission following the galaxy indicating other clumps also producing $\rm Ly\alpha$ emission, the orange hatched circle at the bottom left shows the MUSE PSF. \textit{Right: }Shows the normalized intensity plot of the UVC and $\rm Ly\alpha$ obtained by isophote fitting, which is dominant compared to UVC at all radii and extended twice as far.}
\label{fig:lya_halo}
\end{figure*}

From the continuum subtracted $\rm Ly\alpha$ image of the GSz5BH along with the HST F105W image (we selected this filter because its restframe (1643 \text{\AA} is closest to 1500\text{\AA}), we measured the presence of the $\rm Ly\alpha$ halo. The $\rm Ly\alpha$ image was straightforward as there was no other contamination from the nearby source as they are at a different redshift, and the continuum was removed from the cube (see sec.\ref{subsec:lya_image}). As for the HST image, before the PSF matching, we have to remove all the nearby galaxies from the image; otherwise, when matched with MUSE PSF, which is very large compared to HST, there will be contamination from nearby galaxies. We then masked all the other galaxies using segmentation from the sep module \citep{Barbary2016, Bertin_1996}. We refill these masked pixels with Gaussian background values randomly. As mentioned, MUSE PSF is much larger than HST; we directly convolved the Moffat PSF to the F105W cleaned image. The values for the PSF parameters are calculated from the formula given in the MUSE DRII paper \citep{Bacon2023}. After matching the PSFs, we reprojected the HST image to the MUSE pixel scale ($\rm 0.2^{\prime\prime}/pix$). Now, we compared the normalized isophote intensity profile of the UV continuum and the $\rm Ly\alpha$ as shown in Figure \ref{fig:lya_halo}. The FWHM of the MUSE $\rm Ly\alpha$ is $\sim 0.6^{\prime\prime}$ at the observed wavelength \citep{Bacon2023}, compared to that the $\rm Ly\alpha$ contour ($3\sigma$) extends up to a radius $r_{sma} \sim 1.05^{\prime\prime} (6.33 kpc)$.

We can see the presence of $\rm Ly\alpha$ halo as its profile extends beyond the stellar (UVC) intensity profile. Also, the intensity of the $\rm Ly\alpha$ is brighter even at the smaller radius overall. We fitted the observed isophote intensity profiles with the help of $profiler$ \citep{Ciambur2016}. We used an exponential function convolved with the Moffat PSF \citep{Moffat1969}, $PSF(r) = A[(1+(r/\alpha)^{2})^{-\beta}]$, having FWHM = 0.61$^{\prime\prime}$, $FWHM = 2\alpha\sqrt{2^{1/\beta} - 1}$, and $\beta = 2.8$. Figure \ref{fig:lya_halo} shows the intrinsic exponential profiles. We found the effective radius ($\rm r_{e}$) of $\rm Ly\alpha$ extended twice compared to the UV continuum (stellar) counterpart.

\subsection{Estimating clump velocity}\label{subsec:clump_vel}

We use the MUSE data cube for measuring the peak kinematics of the $\rm Ly\alpha$. We used the $Ly\alpha$ image and corresponding variance image, also provided with the data, of GSz5BH (see section \ref{subsec:lya_image}) and masked the pixels with SNR < 3 in the narrow-band image constructed using MUSE data (see sec. \ref{subsec:lya_image}), SNR is calculated using the corresponding variance image., remaining pixels in the $Ly\alpha$ image are then binned with the Voronoi binning method \citep{Cappellari&Copin2003} with a target SNR = 20. We then use this binning map to bin the cube, resulting in the six binned spaxels. Spectra of the binned $Ly\alpha$ are then extracted from each bin and fitted with a combination of two skewed Gaussian profiles given by following Eq.\ref{eqn:skw_gaussian}, 

\begin{equation}
    f(\lambda) = f_{0}\left(1 + erf\left(\gamma\frac{(\lambda - \lambda_{0})}{\sqrt{2}\sigma}\right)\right)exp\left(\frac{-(\lambda - \lambda_{0})^{2}}{2\sigma^{2}}\right)
    \label{eqn:skw_gaussian}
\end{equation}

\begin{figure*}[ht]
\centering
\includegraphics[width=6in]{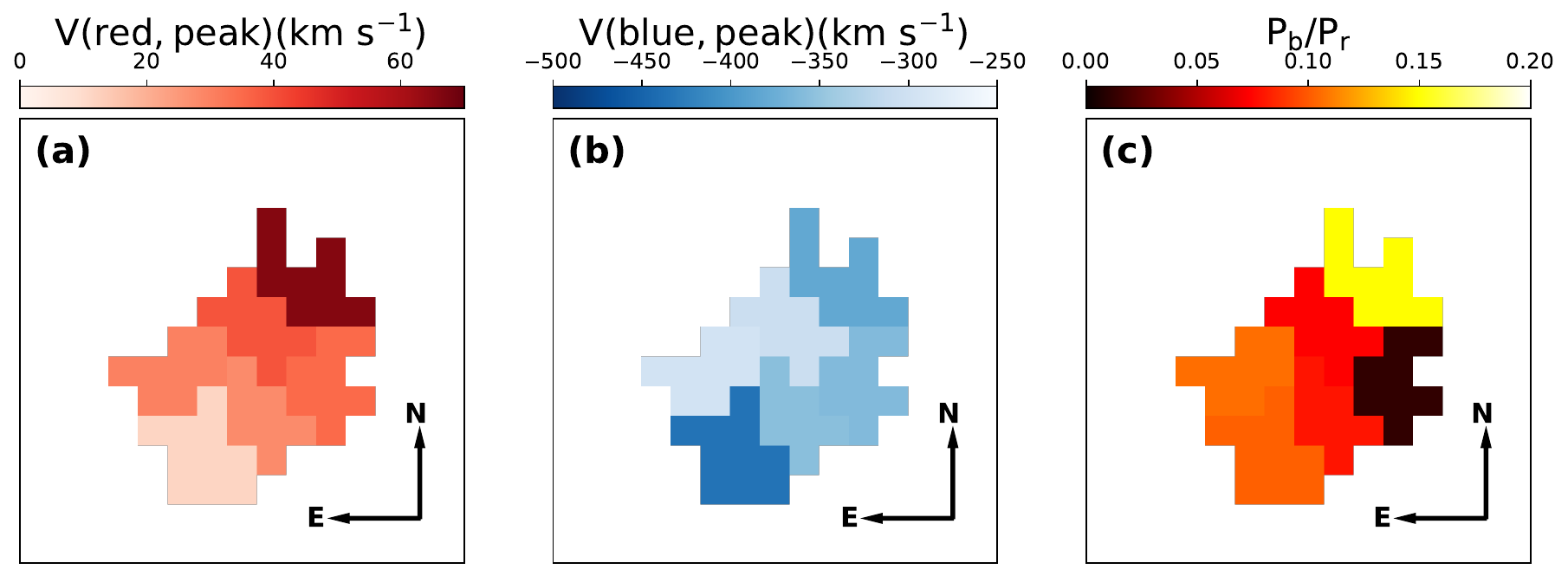}
\includegraphics[clip, trim=0cm 2cm 0cm 3cm, width=6in]{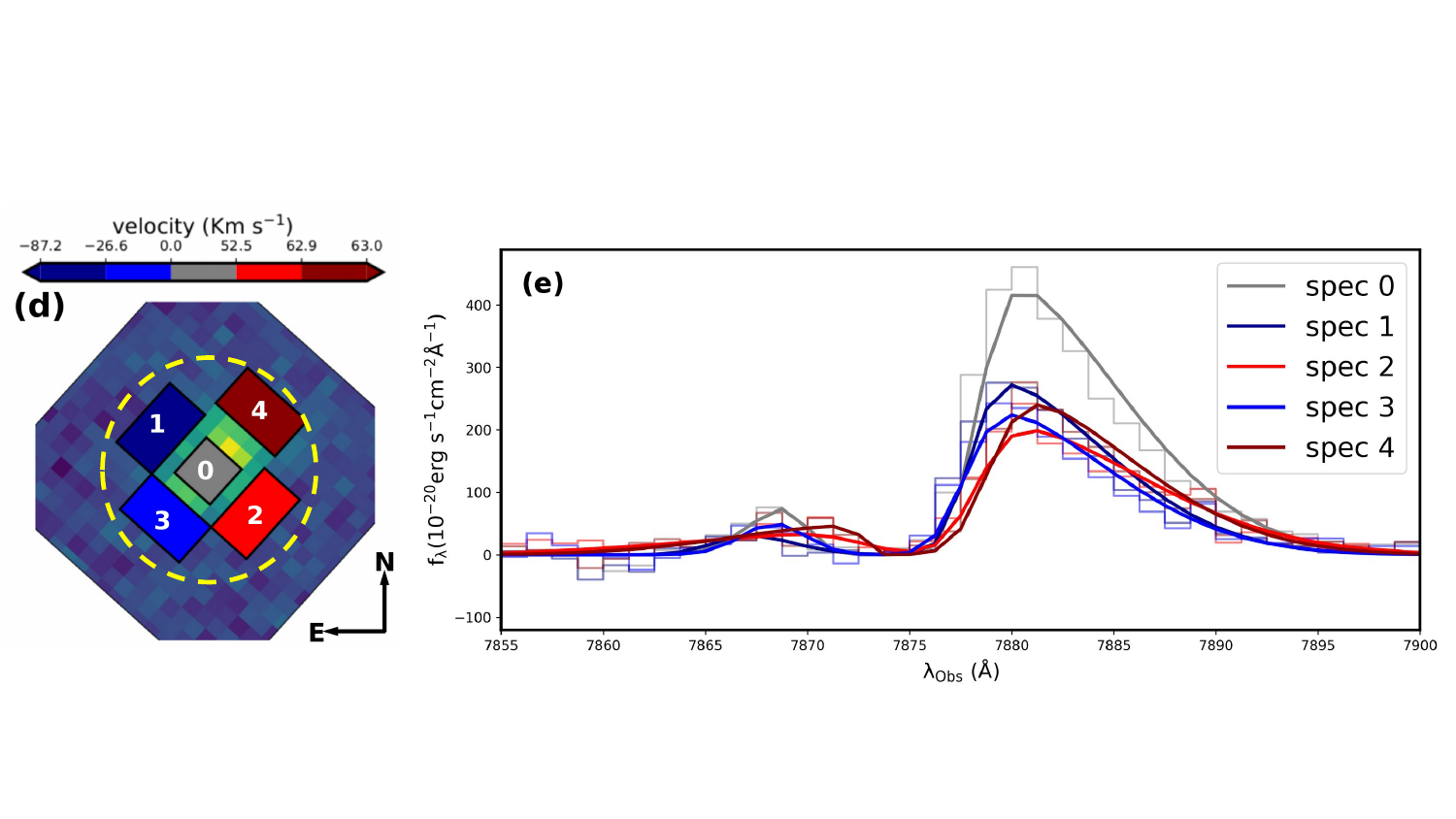}
\caption{\textbf{Kinematics from $\rm \mathbf{Ly\alpha}$: }\textbf{(a)} red peak velocity map of the $Ly\alpha$ emission shows the gradient in velocity along the length, \textbf{(b)} Blue peak velocity map, \textbf{(c)} blue to red peak flux ratio. \textbf{(d)} plot showing the placement of rectangular apertures to extract spectra and the color of apertures indicates calculated velocity \textbf{(e)} shows the asymmetric double peaked Gaussian fit to the extracted spectra from apertures shown in (d) with the same color scheme.}
\label{fig:lya_kin}
\end{figure*}

Where $\gamma$ is the skewness parameter, fitting this equation results in the measurements of peak positions and fluxes. Fitting results are then used to produce red and blue peak kinematic maps with z = 5.48 taken as a systemic redshift. We then create a map of blue to red peak flux ratio (see upper right panel of Fig.\ref{fig:lya_kin}). 

We observe a gradient in the red peak velocity of the GSz5BH, hinting toward the bulk rotation velocity. We then extracted the spectra from the far away region from the center in a rectangular aperture to get an estimate of velocities at the position of clumps, also from the center for reference (see Fig. \ref{fig:lya_kin} bottom panel). We estimate the velocities from line fit by measuring the mean of the red and blue peak positions and comparing them with the center. We observe a rotation w.r.t to the center with the farthest aperture having a velocity of $\sim$ 63 km s$^{-1}$.

To further check the value obtained from the above analysis, we also use the $\rm log\textit{M}_{*}\mbox{--}log\textit{S}_{K}$ scaling relations from the SAMI (Sydney-AAO Multi-object Integral-field spectroscopy) galaxy survey \citep{Barat2019}. Using the following equation, we calculate the $V_{rot}$ of the GSz5BH:

\begin{equation}
logS_{0.5} = a\ logM_{*} + \rm b
\label{eqn:vel_scaling}
\end{equation}

The values for a = 0.39 and b = -1.99 are taken from the best fit for gas dynamics (see Table 2 in \cite{Barat2019}). The K is chosen to be 0.5 for the least scatter in the relation. With the stellar mass $M_{*} = 1.45\times10^{9}\rm M_{\odot}$ of GSz5BH we get an estimate of $V_{rot} \approx $ 44 km s$^{-1}$ assuming $V/\sigma = 2$. 

For the value of the $V_{rot}/\sigma_{g}$, we checked if a value of 2 is reasonable using the following equation from \cite{Simons2017}:

\begin{equation}
    F\left(\frac{V_{\rm rot}}{\sigma_{\rm g}} > x\right) = a\left(\frac{t_{\rm L}}{\rm Gyr} \right) + \rm b
\end{equation}

Where $t_{L}$ is the lookback time in Gyr (which is 12.435Gyr for GSz5BH) for the mass range of $9 < log(M_{*}(\rm M_{\odot})) < 10$, we obtained the fraction of galaxies with $1<V_{rot}/\sigma_{g}<3$ to be $\sim37\%$.

\section{Black Hole growth}\label{sec:black hole growth}

From \cite{Matthee2024}, we have the black hole mass for GSz5BH, measured using the $H\alpha$ broad line component, $M_{BH} = 3.09\times10^{7}\ \rm \rm M_{\odot}$. To get the seed mass of the black hole, we use the following relation from \cite{Volonteri2010}.

\begin{equation}
    T_{\rm growth, BH} = 0.45 \rm Gyr \frac{\epsilon}{1-\epsilon} \lambda^{-1} ln\left(\frac{M_{\rm final}}{M_{\rm initial}}\right)
    \label{eqn:bh_growth}
\end{equation}

We determine the initial seed mass of the black hole using Eqn. \ref{eqn:bh_growth}, assuming a standard radiative efficiency value of $\epsilon = 0.1$ for an accreting Schwarzschild black hole. For the growth period, we assume the black hole formed around the same time as the galaxy, approximately 600 Myr ago. With an Eddington ratio of $\lambda = 1$, the seed mass is estimated at around 190 $\rm M_{\odot}$, which could easily originate from a PopIII star remnant. However, GSz5BH hosts a low-luminosity AGN, and its bolometric luminosity, inferred from broad $H\alpha$ measurements \citep{Greene&Ho2005, Richards2006, Harikane2023}, yields an Eddington ratio of $\lambda = 0.14$. Using this value of $\lambda$, while keeping other parameters unchanged, we calculate a seed mass of $10^{6.76} \rm M_{\odot}$, exceeding the DCBH upper seed mass limit of $10^{6} \rm M_{\odot}$.

As the value of the Eddington ratio is very sensitive for the calculation of the seed mass of the black hole, we took another approach to measure $\lambda$ by using the broadband AGN luminosity in HST/F850LP and JWST/F335M, which corresponds to mean wavelengths of 1400.1 \text{\AA} and 5196.8 \text{\AA} respectively, from \cite{Netzer2019} we use the scaling relation (Eqn. \ref{eqn:l_bolo}) to convert the broadband luminosity into the bolometric luminosity. 

\begin{equation}
    L_{\rm Bol} = c\tilde{\lambda} L_{\rm \tilde{\lambda}}\left(\frac{\tilde{\lambda} L_{\rm \tilde{\lambda}}}{10^{42} \rm erg\ s^{-1}}\right)^{\rm d}
    \label{eqn:l_bolo}
\end{equation}

We choose the F850LP band of the HST/ACS and F335M band of JWST/NIRCAM as they have restframe mean wavelengths that are closest to 1400\text{\AA} and 5100\text{\AA} at the redshift of GSz5BH. We measure the $m_{AB, F850LP} = 27.415$ and $m_{AB, F335M} = 25.07$ of the AGN with the help of PSF modeling as described in \ref{subsec: agn_removal}, which also contains the stellar contribution from the underlying galaxy as we only fit the ePSF at the position of AGN. 

We assume a high dust attenuation value of $A_{v} = 4$ at the position of clump K1 and use the SMC dust attenuation curve with $\rm R_{v} = 2.93$ \citep{Pei1992} as direct measurements are not available. The bolometric luminosity was then calculated using Eqn. \ref{eqn:l_bolo}, with coefficients using c = 7, d = -0.1 for 1400\text{\AA} and c = 40, d = -0.2 for 5100\text{\AA}. This results in bolometric luminosities of 
$L_{Bol} = 4.31\times10^{44}\ \rm{and}\ 3.46\times10^{44}erg\ s^{-1}$, corresponding to Eddington ratios of $\lambda \sim 0.1\ \rm{and}\ 0.09$
for both filters, which are in close agreement with each other. This calculation also does not exclude the contribution of underlying stellar light. While we selected a high $A_{v}$, a lower value would further reduce the Eddington ratio, making the current estimate of $\lambda = 0.14$ very conservative.

With the value Eddington ratio $\lambda = 0.14$, $T_{\rm growth, BH} = 0.6\ \rm Gyr$ and $\epsilon = 0.1$, the required seed mass of the black hole is $M_{seed} = 10^{6.8}\rm M_{\odot}$ which is $\sim 6$ times higher than the DCBH upper limit of the seed mass. Assuming the black hole formed around 1 Gyr ago, which corresponds to the age of the universe at this redshift (z = 5.48), and keeping all other parameters unchanged, we would still require a seed black hole with a mass of $M_{seed} = 10^{6.2}\rm M_{\odot}$ given the Eddington ratio of $\lambda = 0.14$. Hence, the accretion rate of the GSz5BH black hole must have been higher in the past, so to explain the very high observed black hole mass ($\sim$2.1\% of the host galaxy stellar mass compared to the local universe with value $\sim$0.1\%), we need a mechanism that can efficiently feed the gas in the innermost regions of the galaxy.

\subsection{Clump migration timescales} \label{subsec:clump_migration_time}

In the previous section \ref{sec:black hole growth}, it was clear that the current value of the Eddington ratio cannot explain the observed mass of the black hole in GSz5BH, which is the case with most high-z galaxies. With GSz5BH having big star-forming clumps, which are a significant fraction of its total stellar mass, they can be easily dragged into the center of the galaxy (assuming the black hole hosting clump (K1) is being the center of the gravitational potential). Torques on such massive clumps generated by the dynamical friction from the dark matter halo alone can easily make them inspiral towards the center. This clump accretion can fuel the growth of the black hole and feed the AGN.  

To calculate the timescale of clumps in the galaxy, we assumed the clump containing the AGN (K1) to be the center. We calculated using the logarithmic spherical potential of the dark matter:

\begin{equation}
    \Phi(r) = \frac{V_{0}}{2} ln[R_{\rm c}^{2} + r^{2}]
    \label{eqn:dm_potential}
\end{equation}

Using Poisson's equation to find the mass distribution from the assumed dark matter potential given in equation \ref{eqn:dm_potential}:

Using Poisson's equation above, we obtained the density distribution for the assumed dark matter potential and the mass distribution.

As the above equation shows, the dark matter mass doesn't converge. The $\rm M_{dyn}(r)$ is calculated using the equation below. 

\begin{equation}
    M_{\rm dyn}(r) = \frac{V_{0}^{2}\ r}{G} = M_{*} + M_{dm}(r)
    \label{eqn:M_dyn}
\end{equation}

Where we use the value of $\ V_{0} = 62.8$ km s$^{-1}$ calculated at a distance of r $\sim$ 4.8 kpc relative to the center, which is farther away from the farthest clump as the distance between K1 \& K3 is 2.86 kpc hence we assume it to be the saturation velocity, from this calculation, we get the value of $\rm M_{dyn}(r) = 4.43\times10^{9}\rm M_{\odot}$. We then use this value of the $\rm M_{dyn}$ to calculate the core radius of the dark matter using the following equation.

\begin{equation}
    R_{\rm c} = R_{\rm out}\sqrt{\left(\frac{M_{\rm dyn}}{M_{*}} - 1\right)}
    \label{eqn:rc}
\end{equation}

From the Eqn. \ref{eqn:rc}, we obtained a value of the core radius of $R_{c} = 6.85\ kpc$. We then used this value to calculate the inspiral timescale of the clump K3 falling into the central region (K1) due to dynamical friction of dark matter only, using the following expression also used in \cite{Borgohain2022}. Eqn:\ref{eqn:t_insp} assumes the stellar clumps to be bound self-gravitating structures embedded in the surrounding dark matter particles, we assume that clumps don't lose mass while they inspiral towards the center. The effect of gas interaction is not taken into account in this equation, and it treats the velocity distribution of dark matter particles as Maxwellian \citep{Binney&Tremaine2008, Elmegreen2012}.

\begin{equation}
    \frac{T_{\rm inspiral}}{T_{\rm orbital}} = \frac{\alpha M_{*}/M_{\rm c}}{ln(\alpha M{*}/M_{\rm c})}\frac{X_{\rm dm}^{2}}{\xi(X_{\rm dm})}\frac{I_{\rm dm}(R_{\rm c}, R_{\rm out}, R_{\rm in})}{2\pi(R_{\rm out}/R_{\rm c})}
    \label{eqn:t_insp}
\end{equation}

Where, 

\begin{equation}
    T_{\rm orbital} = \frac{2\pi R_{\rm out}^{3/2}}{\sqrt{G\alpha M_{*}}}
    \label{eqn:t_orb}
\end{equation}

and

\begin{equation}
    I_{dm} = \int_{\frac{R_{out}}{R_{c}}}^{\frac{R_{in}}{R_{c}}} \frac{2+x^{2}}{3+x^{2}}\frac{x^{2}}{\sqrt{1+x^{2}}} \,dx
    \label{eqn:Idm}
\end{equation}

The indefinite solution of the above integral has the form, 

\begin{multline}
    I_{dm}(x) = \frac{1}{2} \left( x\sqrt{x^2+1}\ +3 \ln
   \left(\sqrt{x^2+1}-x\right)\right)+\\ 
   \sqrt{\frac{3}{2}} \tanh^{-1}\left(\frac{x^2-x\sqrt{x^2+1} +3}{\sqrt{6}}\right)
   \label{eqn:Idm_sol}
\end{multline}

\begin{figure}[ht]
\nolinenumbers
\centering
\rotatebox{0}{\includegraphics[width=0.5\textwidth]{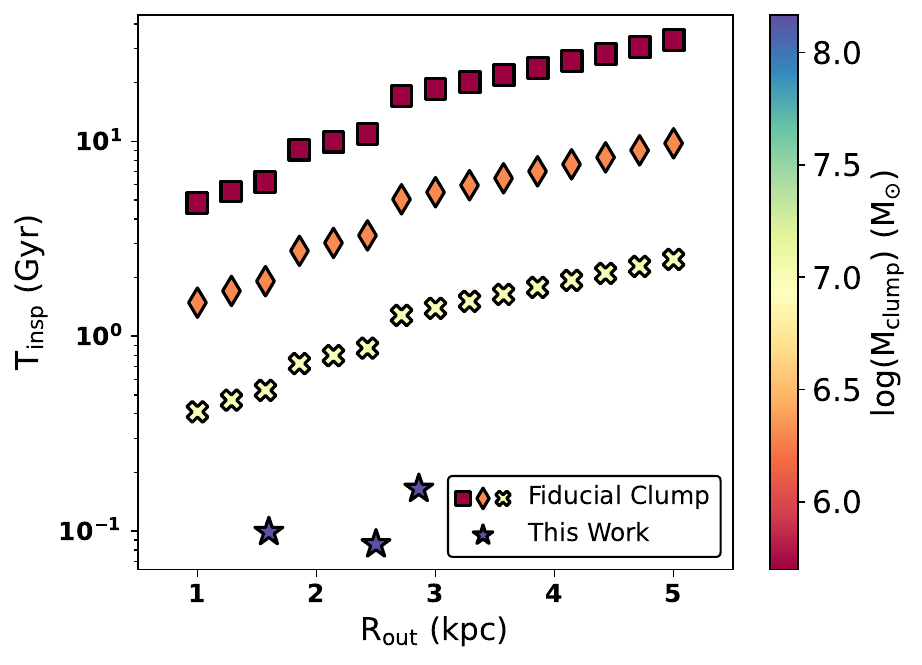}}
\caption{Comparison of inspiral timescale vs distance ($\rm R_{out}$) from the centre for different clump masses. The color bar indicates the range of clump masses used in this plot. The symbols - square, diamond, and cross represent test clumps, with the highest mass of the test clumps being $10^{7}M{\odot}$. The star symbols represent actual clumps detected in the galaxy.}
\label{fig:clump_acc_plot}
\end{figure}

In the equation \ref{eqn:t_orb}, we use $R_{in} = 0\ {\rm kpc} $ and $\alpha = 3$, where $R_{in}$ is the radius where the clumps coalesce with the center $\alpha$ is the ratio of dynamical to stellar mass, we determine an inspiral timescale of 0.16 Gyr for the clump K3. Similarly, we calculated $T_{inspiral}$ for the clump C and K2 (at a projected distance of 2.5 and 1.6 kpc) with $\alpha = 3$, which turned out to be 0.09 and 0.1 Gyr, respectively (see Table \ref{tab:insp_time}). Using the above calculations of inspiral time scales and stellar masses of each clump, we calculate the clump accretion timescale for K1 using the following equation.

\begin{equation}
    \dot{M}_{\rm clump} = \sum_{i=1}^{N} \frac{M_{\rm clump, i}}{T_{\rm inspiral, i}}
\end{equation}

The total clump accretion rate based on the above equation is estimated to be $\dot{M}_{\rm clump} \sim 14\rm M_{\odot}yr^{-1}$.

Simulation studies indicate that the dynamical friction process does not completely eliminate orbital angular momentum; instead, it leads to the formation of a circumnuclear disk of roughly 100 pc in size \citep{Hopkins&Quataert2010}. However, by changing $R_{in} = 0.1\ kpc$ instead of 0 kpc, the inspiral timescales are found to modify insignificantly. It is noted that the above equation for inspiral timescale calculation does not capture the movement of material in the sub-kpc region. Nevertheless, Eqn \ref{eqn:t_insp} is more sensitive to $R_{out}$ and clump mass, we estimate inspiral timescales for a range of clump mass and initial location of the clumps in the galaxy, see Fig.~\ref{fig:clump_acc_plot}.  The lower mass clumps (typically million solar mass) would take more than a few billion years while the higher mass (typically 100 million solar mass) ones would need less than a billion years had they started from the same initial location in the galaxy. Note that in all cases, the clumps would migrate to the circumnuclear disk at the end. Subsequent accretion of material from the circumnuclear disk to the black hole is hard but not as hard as the final parsec problem \citep{Milosavljevic&Merritt2003}. The material in the circumnuclear disk can lose angular momentum by different physical processes, including gas dynamical and stellar dynamical processes. One possibility is the formation of a non-axisymmetric pattern such as a nuclear bar or nuclear spiral in the circumnuclear disk \citep{Combes2003} and aid the material to lose angular momentum further. However, a detailed treatment of this is beyond the scope of this paper. Interestingly, once the material is within the black hole sphere of influence \citep{Merritt&Ferrarese2001} whose radius is given by $R_{influ} = GM_{BH}/\sigma^{2}$ where $\sigma$ refers to the velocity dispersion, it is likely to be accreted by the black hole. For the pseudo-isothermal halo assumed in our calculation, the value of $\sigma = V_{max}/\sqrt{2} = 44$~kms$^{-1}$ and $R_{influ}$ turns out to be $\approx 67\ pc$. Since this is smaller than the assumed radius of the circumnuclear disk, the calculated mass accretion rate is more representative of inflow into the inner regions rather than direct accretion onto the black hole.

\subsection{Inspiral model of black hole growth} \label{subsec:insp_model}

Clump accretion is discrete as the clumps form and migrate toward the center via dynamical friction. Currently, three detected bright stellar clumps (C, K2, and K3 apart from K1) are in the GSz5BH, each containing a significant fraction of the total stellar mass. The clump migration times are of the order of $\sim 100 ~\rm Myrs$ as calculated in the section \ref{subsec:clump_migration_time}. Assuming such clumps were present in the past. New ones will be forming in the future due to the abundance of the HI gas (as probed by the $\rm Ly\alpha$ emission in section \ref{subsec:lya_halo}). Due to short migration timescales, we assume a constant clump accretion model with $\dot{M} = \eta \dot{m}_{{\rm clump}}$ by integrating this equation from we obtain:

\begin{equation}
    M_{\rm BH}(t) = M_{\rm seed} + \eta \dot{m}_{\rm clump}\rm t
    \label{eqn:clump_accretion}
\end{equation}

In the above equation \ref{eqn:clump_accretion}, $\eta$ represents the efficiency of the accretion, i.e., what accreted mass went into feeding the black hole, $M_{\rm seed}$ is the seed mass at the beginning of the accretion. In Figure \ref{fig:bh_growth}, we show the clump accretion by the hatched region governed by this equation (see Eqn.\ref{eqn:clump_accretion}) with band representing the range $\eta = 0.001\ {\rm and}\ \eta = 0.01$, i.e., feeding efficiency of 0.1 and 1\%. With only 1\% of feeding efficiency, we can drive enough matter into the black hole to grow it to the observed mass. So, this episodic clump accretion seems responsible for inflating the Eddington ratio in the past via driving gas and stars to the innermost regions to achieve such supermassive black hole mass in such a short time (with the age of the universe only being <1Gyr at this redshift).

\section{Conclusion \& Discussion}

In this work, we tested the idea of the clump-assisted rapid growth of the central supermassive black hole in a high redshift galaxy having a 30 million solar mass black hole. We showed that even with 1\% of the feeding efficiency, we can inspiral the matter in the galaxy's central region with a timescale of $\sim$100Myr. With the help of only the dynamical friction from the dark matter halo, we achieved this timescale. Including the gas dynamical friction and clump-clump interaction will further reduce these times. Hence, the idea of growth via the clump's accretion seems to work within the GSz5BH framework, and this model is viable in this galaxy's context, along with the other growth models discussed in the introduction. 

\section*{Acknowledgments}

We sincerely thank Debra Elmegreen for the valuable ideas she shared during the project discussions. We thank Frederic Bournaud for his constructive and insightful comments, which significantly enhanced the clarity of this study.  Additionally, we acknowledge the Mikulski Archive for Space Telescopes (MAST) at the Space Telescope Science Institute (STScI) for maintaining and providing public access to the HST and JWST imaging datasets, which are extensively used in this research. We thank the IUCAA HPC (High-performance Computing) Pegasus server for providing computational facilities.

\bigskip

\noindent {\bf Data availability}\\
The HST imaging data are available at \cite{https://doi.org/10.17909/t91019}, The JADES/JWST survey data is found here \cite{https://doi.org/10.17909/8tdj-8n28}. MUSE data used in this paper can be accessed from \cite{Bacon2017}.

\noindent {\bf Code availability}\\
We have used standard data reduction tools in Python ({\url{https://www.python.org/}}), and the publicly available code SExtractor ({\url{https://www.astromatic.net/software/sextractor}} for this study. This research made use of Astropy \citep{Astropy2013, Astropy2018},  photutils ({\url{https://photutils.readthedocs.io/en/stable/}}), which are community-developed core Python packages for Astronomy, PyPHER ({\url{https://pypher.readthedocs.io/en/latest/}}), we use Ned Wright's web-based cosmology calculator \citep{Wright2006} for distance calculations in this paper {\url{https://www.astro.ucla.edu/%7Ewright/CosmoCalc.html}},  PROFILER ({\url{https://github.com/BogdanCiambur/PROFILER}}), ds9 \citep{SAO2000, Joye&Mandel2003},
CIGALE ({\url{https://cigale.lam.fr/}}) The MPDAF for extracting MUSE spectra \citep{Bacon2016, Piqueras2019}, Mathematica \citep{Mathematica}

\appendix

\section{Appendix information}

\begin{figure*}[ht]
\centering
\includegraphics[width=1.0\textwidth]{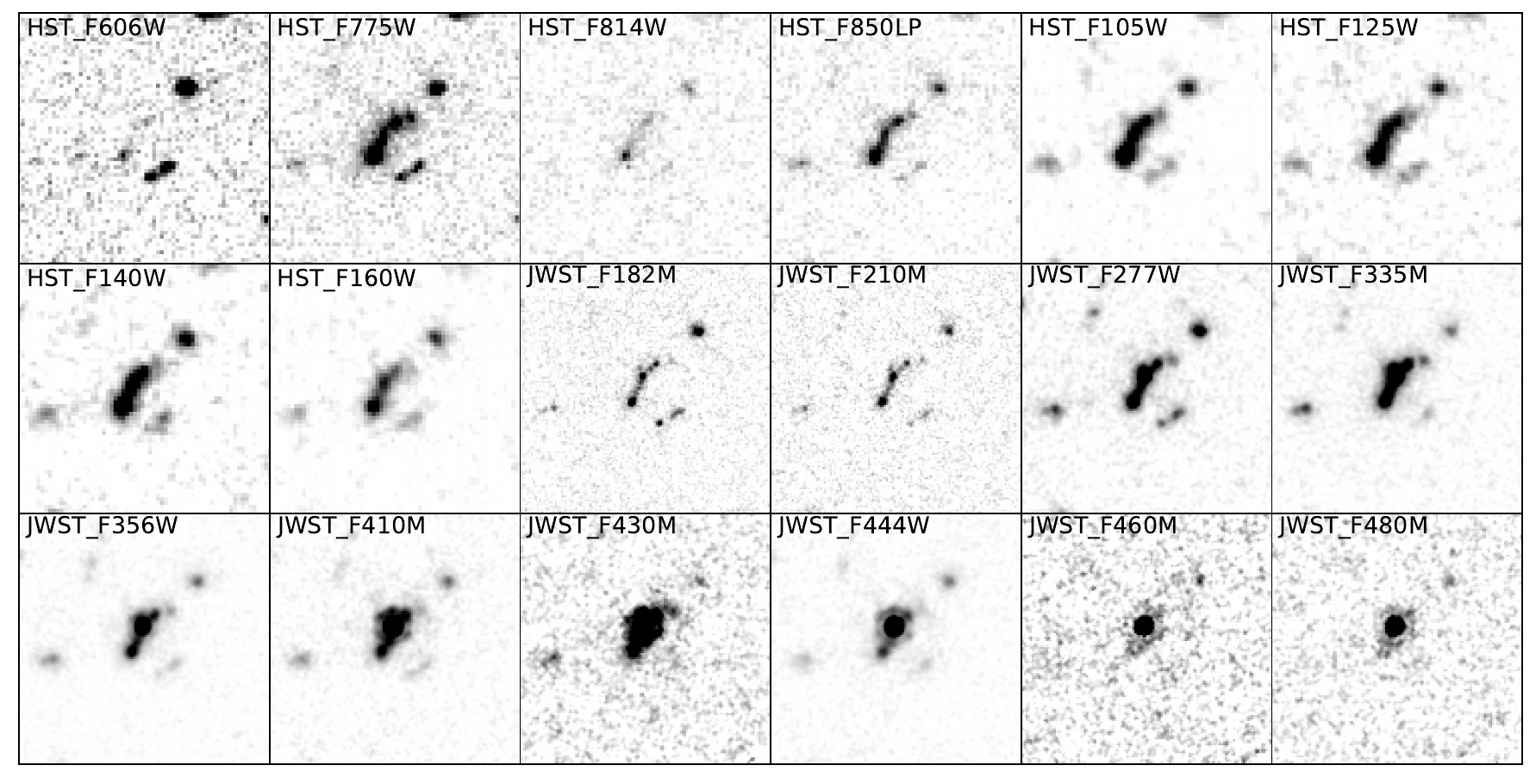}
\caption{\textbf{Postage stamp band images of GSz5BH:} HST (UV) and JWST (UV, Optical) restframe band images. PSF-like feature (AGN) starts appearing from the JWST F277W band, probing the restframe optical.}
\label{fig:hst_jwst_cutouts}
\end{figure*}

\begin{figure*}[ht]
\centering
\includegraphics[width=1.0\textwidth]{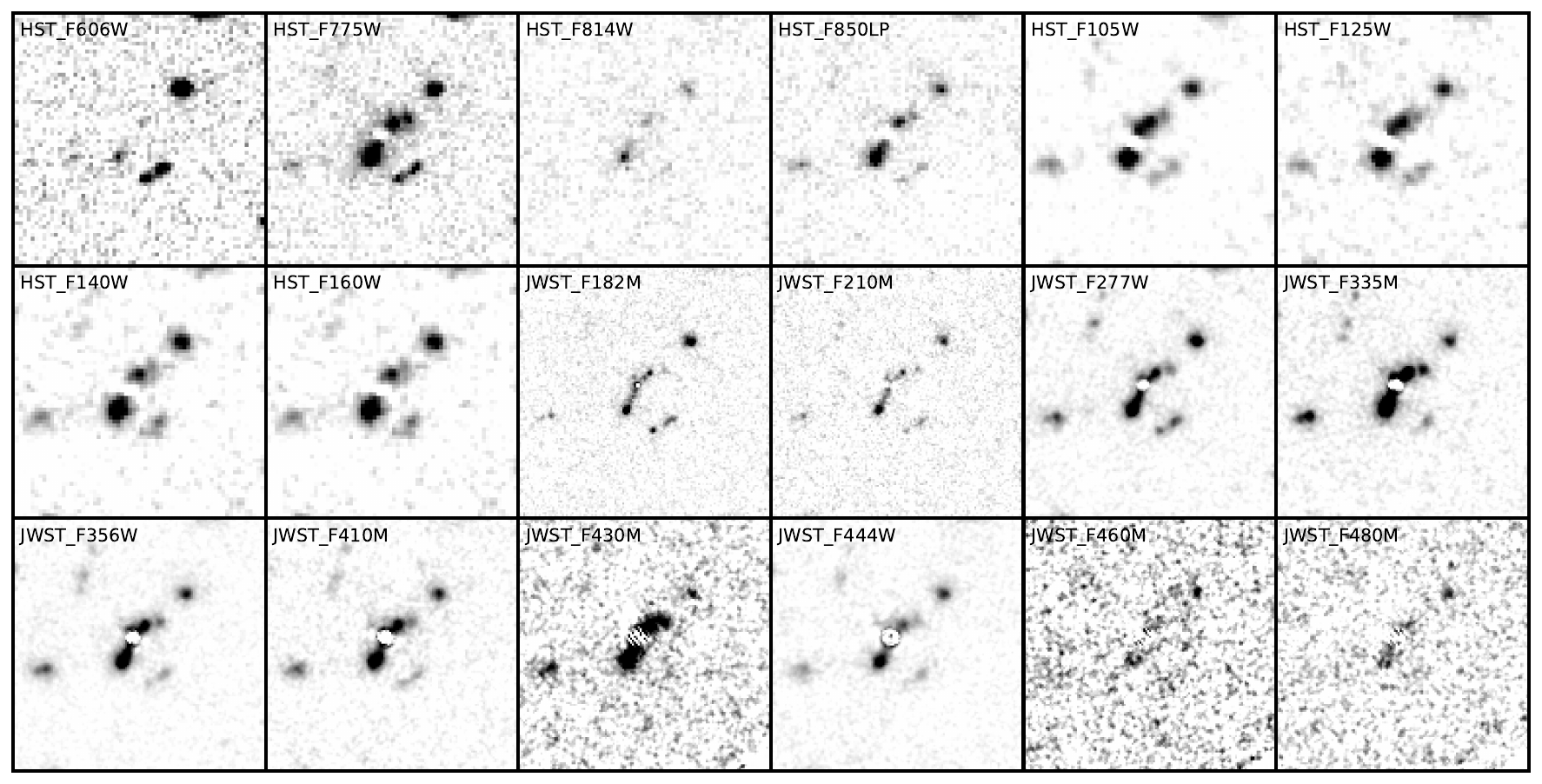}
\caption{\textbf{GSz5BH images after AGN subtraction:} HST (UV) and JWST (UV, Optical) restframe band images after removal of AGN light via PSF modeling.}
\label{fig:hst_jwst_cutouts_agn_sub}
\end{figure*}

\begin{figure*}[ht]
\centering
\includegraphics[width=6in]{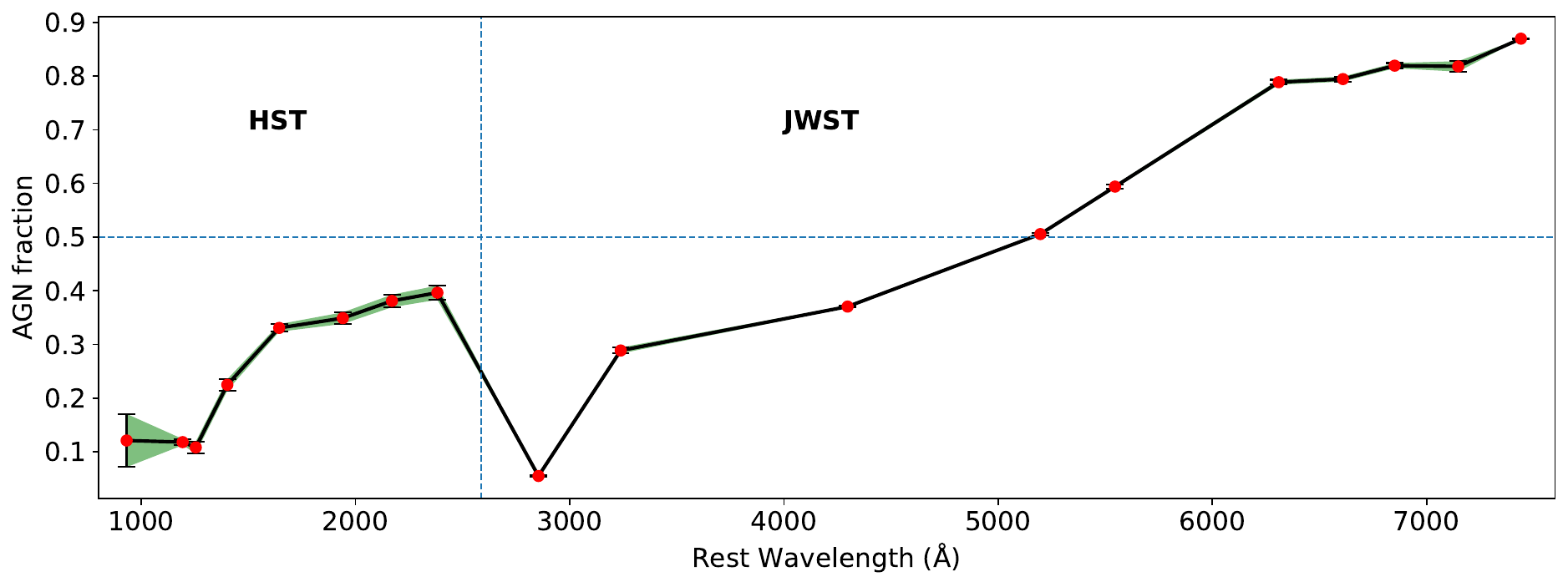}
\caption{\textbf{AGN fraction as a function of wavelength: }Ratio of the AGN light measured from the PSF modeling to the total light of the galaxy GSz5BH w.r.t the wavelength, the vertical dotted line separates the HST and JWST bands. In the UV bands and the first optical band (F277W), the AGN fraction is < 0.5. The sudden dip in the first UV band of JWST is due to systematic effects of increased resolution and faintness of the AGN in UV.}
\label{fig:agn_fraction}
\end{figure*}

\begin{table*}[ht]
    \renewcommand{\arraystretch}{1.5}
    \setlength{\tabcolsep}{20pt}
    \centering
    \begin{tabular}{|c|c|c|c|c|}
    \hline
         \textbf{Observed band} & \textbf{Full galaxy}  & \textbf{C} & \textbf{K2} & \textbf{K3} \\
          & $m_{AB}$ & $m_{AB}$ & $m_{AB}$ & $m_{AB}$ \\
    \hline
         HST\_F606W & $28.91\pm0.12$ & - & - & - \\
         HST\_F775W & $26.25\pm0.02$ & - & - & - \\
         HST\_F814W & $26.05\pm0.01$ & - & - & - \\
         HST\_F850LP & $26.09\pm0.04$ & - & - & - \\
         HST\_F105W & $26.48\pm0.01$ & - & - & - \\
         HST\_F125W & $26.50\pm0.02$ & - & - & - \\
         HST\_F140W & $26.63\pm0.01$ & - & - & - \\
         HST\_F160W & $26.80\pm0.03$ & - & - & - \\
         JWST\_F182M & $25.94\pm0.002$ & $26.72\pm0.007$ & $27.21\pm0.011$ & $29.48\pm0.06$ \\
         JWST\_F210M & $26.52\pm0.003$ & $26.90\pm0.008$ & $27.74\pm0.017$ & $29.48\pm0.06$ \\
         JWST\_F277W & $26.08\pm0.002$ & $26.81\pm0.007$ & $27.57\pm0.014$ & $28.99\pm0.04$ \\
         JWST\_F335M & $25.09\pm0.001$ & $25.80\pm0.003$ & $26.49\pm0.005$ & $27.75\pm0.01$ \\
         JWST\_F356W & $25.69\pm0.001$ & $26.34\pm0.004$ & $27.01\pm0.008$ & $28.40\pm0.02$ \\
         JWST\_F410M & $25.69\pm0.001$ & $26.35\pm0.005$ & $27.07\pm0.009$ & $28.25\pm0.02$ \\
         JWST\_F430M & $25.14\pm0.001$ & $25.76\pm0.003$ & $26.50\pm0.006$ & $27.41\pm0.01$ \\
         JWST\_F444W & $26.17\pm0.002$ & $26.77\pm0.007$ & $27.57\pm0.014$ & $28.70\pm0.03$ \\
         JWST\_F460M & $26.48\pm0.003$ & $27.23\pm0.011$ & $27.91\pm0.02$ & $29.15\pm0.05$ \\
         JWST\_F480M & $26.89\pm0.004$ & $27.11\pm0.009$ & $27.84\pm0.019$ & - \\
    \hline
    \end{tabular}
    \caption{\textbf{GSz5BH AB magnitudes: }Observed magnitudes of Full galaxy and the individual clumps with aperture and foreground dust correction.}
    \label{tab:mag}
\end{table*}

\begin{table*}[ht]
    \centering
    \begin{tabular}{|l|l|l|l|l|}
    \hline
         properties/best-fit value & full galaxy (without K1) & C & K2 & K3 \\
    \hline
         $\rm \chi_{r}^{2}$ & 3.9 & 0.32 & 1.2 & 0.4\\
         E(B-V) & 0 & 0 & 0 & 0.078\\
         $\rm E(B-V)_{factor}$ & 0.44 & 0.44 & 0.44 & 0.6\\
         $\rm f_{esc}$ & 0.1 & 0.1 & 0 & 0\\
         $\rm f_{dust}$ & 0 & 0.4 & 0.4 & 0\\
         $\rm \beta -slope$ & -2.64 & -2.88 & -2.81 & -1.93\\
         $\rm Z_{gas}$ & 0.003 & 0.003 & 0.002 & 0.006\\
         $\rm Z_{*}$ & 0.004 & 0.004 & 0.0004 & 0.0001\\
         $\rm SFR\_100 (M_{*}yr^{-1})$ & $1.17\pm0.16$ & $1.2\pm 0.4$ & $0.6\pm 0.2$ & $0.11\pm 0.03$\\
         $\rm M_{*} (\rm M_{\odot})$ & $\rm (1.22\pm0.21)\times10^{9}$ & $\rm (8.05\pm2.7)\times10^{8}$ & $\rm (4.18\pm1.4)\times10^{8}$ & $\rm (1.47\pm0.44)\times10^{8}$\\
         Age (Myr) & $588.\pm142.2$ & $618.8\pm278.6$ & $637.2\pm274.8$ & $602.4\pm250.7$\\
         $\rm D\_4000$ & $0.45\pm0.03$ & $0.58\pm0.03$ & $0.58\pm0.03$ & $0.48\pm0.04$\\
    \hline
    \end{tabular}
    \caption{\textbf{SED modeling: }Best-fit parameters of total and individual clump SED of GSz5BH with all bands (UV to optical) for total galaxy and only JWST bands for clumps.}
    \label{tab:best_fit}
\end{table*}

\begin{table*}[ht]
    \centering
    \begin{tabular}{|c|c|c|c|}
    \hline
         clump & $\rm D_{K1} (kpc)$ & $\rm T_{insp} (Gyr)$ & $\rm T_{orbital} (Gyr)$ \\
    \hline
         C & 2.50 & 0.09 & 0.30\\
         K2 & 1.60 & 0.10 & 0.22\\
         K3 & 2.86 & 0.16 & 0.23\\
         
    \hline
    \end{tabular}
    \caption{\textbf{Clump migration timescales: } Observed and calculated parameters for clumps in the GSz5BH $\rm D_{K1}$ is the projected distance of a clump from the center K1, while $\rm T_{insp} (Gyr)$ and $\rm T_{orbital} (Gyr)$ are inspiral and orbital time.}
    \label{tab:insp_time}
\end{table*}

\clearpage
\bibliographystyle{aasjournal}
\bibliography{citations.bib}

\end{document}